\documentclass[showpacs,aps,prl,preprint,superscriptaddress]{revtex4}

\usepackage{graphicx}
\usepackage{longtable}
\usepackage{multirow}

\begin{document}

\title{Defect induced rigidity enhancement in layered semiconductors}

\author{Zs. Rak}
%\email[email:]{rak@pa.msu.edu}
\author{S. D. Mahanti}
\affiliation{Department of Physics and Astronomy, Michigan State University, East Lansing, MI 48824}
\author{K. C. Mandal}
\affiliation{EIC Laboratories, Inc, 111 Downey Street, Norwood, Massachusetts 02062}
\author{N. C. Fernelius}
\affiliation{AFRL/RX, WPAFB, OH 45433}

\date{\today}

\begin{abstract}
We discuss the mechanism responsible for the observed improvement in the structural properties of In doped GaSe, a layered material of great current interest. Formation energy calculations show that by tuning the Fermi energy, In can substitute for Ga or can go as an interstitial charged defect$\left( \text{In}_{\text{i}}^{\text{3+}} \right)$. We find that $\text{In}_{\text{i}}^{\text{3+}}$ dramatically increases the shear stiffness of GaSe, explaining the observed enhancement in the rigidity of In doped p-GaSe. The mechanism responsible for rigidity enhancement discussed here is quite general and applicable to a large class of layered solids with weak interlayer bonding.
\end{abstract}

% PACS

%\keywords{}

\maketitle

Defects can dramatically alter the physical properties of solids in general, semiconductors in particular. Whereas the electronic transport properties of semiconductors are controlled by defects (doping and scattering), their mechanical properties are less so. In this letter we show that the mechanical properties of GaSe, a layered semiconductor of great current interest, can be controlled by suitable choice of defects and the Fermi energy. The idea is very general and can be applied to a large class of soft layered solids.

GaSe is one of the members of a class of semiconducting compounds formed out of group III (Al, Ga, In) and group VI (Se, Te) elements. These are quasi-two dimensional layered systems containing blocks consisting of four atomic planes (Se-Ga-Ga-Se). The atoms inside a block are strongly bonded (covalent and ionic) whereas the interaction between different blocks is weak Van der Waals (VdW) type. GaSe is a highly efficient nonlinear optical material, with applications in second harmonic generation, frequency mixing, and generation/detection of terahertz radiation~\cite{frene;pcgcm94, dimit;springer99}. However due to the weak VdW interaction between the blocks its mechanical properties such as hardness and cleavability are not satisfactory for practical applications. It has nearly zero hardness by Mosh scale, cleaves easily along the planes parallel to the atomic layers and the nonlinear properties are difficult to reproduce from sample to sample~\cite{frene;pcgcm94}. These difficulties hamper the use of large GaSe crystals in practical applications. Dramatic improvement in the crystal quality has been reported in GaSe doped with $0.1-3$ mass\% In~\cite{voevo;opmat04}. Since In and Ga are isovalent, why In doping increases the rigidity of GaSe is a big puzzle. In this letter we discuss the results of ab initio electronic structure calculations involving In defects, explaining this puzzle. The fundamental question is where do the In impurities go in the host lattice and whether or not they strongly affect the local bonding between both host-host and host-defect pairs leading to the hardening of the elastic constants. We show that by a suitable control of the location and the charge state of the In defect one can indeed dramatically enhance the interlayer rigidity of GaSe.

\begin{figure}[tbp]
\begin{center}$\,$
\begin{tabular}{c c}
\includegraphics[width=1.6875in,keepaspectratio=1]{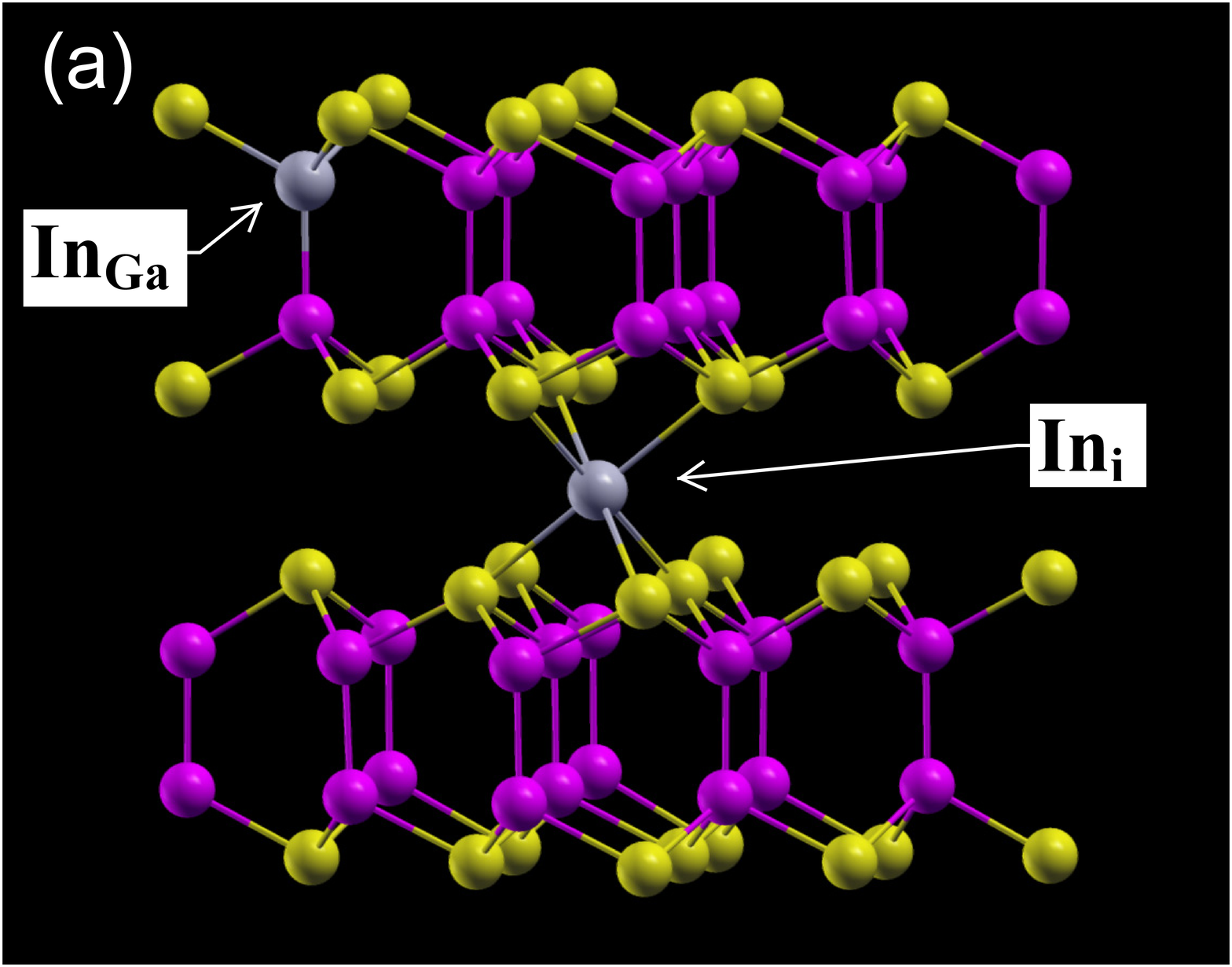} &
\includegraphics[width=1.6875in,keepaspectratio=1]{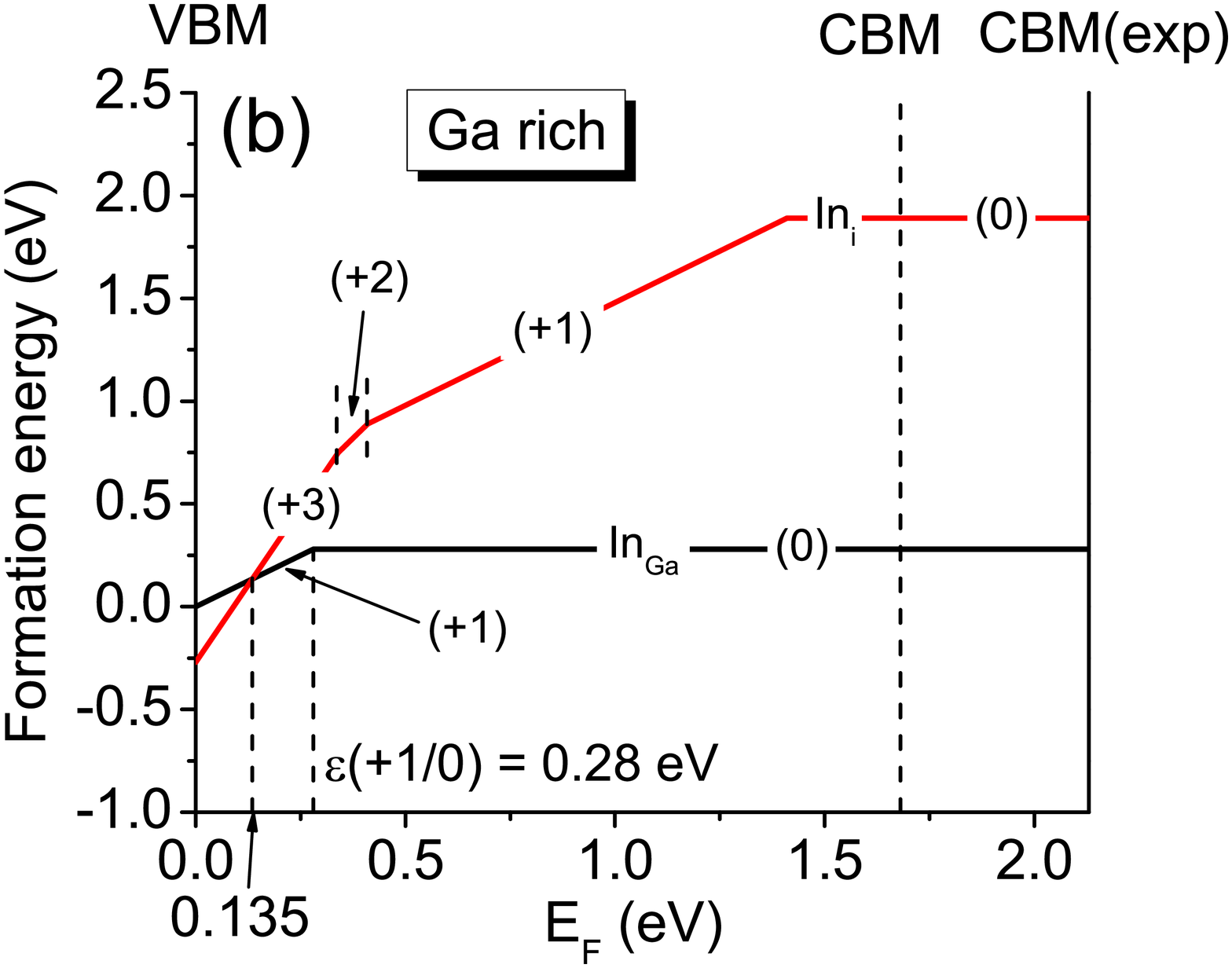} \\
\end{tabular}
\end{center}
\caption {(a) The crystal structure of layered GaSe. The interstitial site is located between the GaSe layers, equally far from the 6 NN Se atoms; (b) Formation energies associated with In$_{\text{Ga}}$ and In$_{\text{i}}$. When $E_{F}$ is close to VBM (p-type GaSe), the In interstitial impurity becomes stable.}
\protect\label{fig;struct_form_ene}
\end{figure}

Due to its layered structure GaSe crystallizes in different polytypes. Since the physical properties of the different polytypes are quite similar we have chosen $\beta-\text{GaSe}$  as the test case. $\beta-\text{GaSe}$ crystallizes in layered, hexagonal structure, having the space group $P6_{3}/mmc$. As mentioned before GaSe consist of blocks of four atomic planes; two planes of Ga atoms sandwiched between two planes of Se atoms as illustrated in Fig.~\ref{fig;struct_form_ene}(a). In each plane the Ga or Se atoms are arranged in a two-dimensional hexagonal lattice. An important feature of this compound is the existence of Ga-Ga dimers oriented perpendicular to the layers; their presence leads to different local geometry of Ga and Se.  Ga has three Se nearest neighbors (NN) and one Ga NN, while each Se is bonded to three Ga atoms and no Se atoms. The presence of Ga-Ga dimers leads to bonding and antibonding Ga $4s$ states. The antibonding $4s$ strongly hybridizes with the Se $p$ bands and opens up a gap near the Fermi energy, leading to a semiconductor~\cite{rak;jpcs09}. It is interesting to note that two formula units $(\text{Ga}_{2}\text{Se}_{2})$ satisfy the 18 electron rule (6 from 2 Ga and 12 from 2 Se) which is conducive to gap formation as in half-Heusler systems~\cite{larso;prb99}. The unit cell of $\beta-\text{GaSe}$ however consists of 4 formula units containing 2 Ga dimers.

Ab initio electronic structure calculations were carried out using the projector-augmented-wave (PAW)~\cite{kress;prb99} method, within density functional theory (DFT) as implemented in the Vienna Ab-initio Simulation Package (VASP)~\cite{kress;prb94,kress;prb96}. The exchange-correlation potential was approximated by the Ceperley-Adler local density approximation (LDA)~\cite{ceper;prl80}. This exchange-correlation potential was chosen over the gradient corrected version (GGA) because it is known that GGA underestimates the binding energies, which results in an overestimation of the lattice parameters~\cite{ander;prl96}. Since the interaction between the blocks of GaSe is week VdW type, the ``GGA effect'' becomes much more significant in the direction perpendicular to the atomic layers, resulting in a theoretical structure that is overly elongated along the crystallographic $c$-axis~\cite{rak;jpcs09}. Such an elongated structure would not be appropriate for a theoretical investigation of the elastic properties of GaSe.  In all calculations the outer $s$, $p$, $d$ orbitals of the Ga and In atoms as well as the $s$ and $p$ orbitals of the Se were included in the valence states, while the rest were treated as core states. The cut-off energy for the plane wave basis was set to 300 eV and the convergence of self-consistent cycles was assumed when the energy difference between them was less than $10^{-4}$ eV.
%
%%%%%%%%%%%%%%%%%%%%%%%%%%%%%%%%%%%%%%%%%%%%%%%%%%%%%%%%%%%%%%%%
%   FIGURE
%%%%%%%%%%%%%%%%%%%%%%%%%%%%%%%%%%%%%%%%%%%%%%%%%%%%%%%%%%%%%%%%

%
%%%%%%%%%%%%%%%%%%%%%%%%%%%%%%%%%%%%%%%%%%%%%%%%%%%%%%%%%%%%%%%%
%   FIGURE
%%%%%%%%%%%%%%%%%%%%%%%%%%%%%%%%%%%%%%%%%%%%%%%%%%%%%%%%%%%%%%%%
%
Various defect properties were investigated using a 3x3x1 supercell model containing 72 atoms (36 Ga and 36 Se) (see Fig.~\ref{fig;struct_form_ene}(a)). Two types of In defects were studied, (i) In substituting for Ga (In$_{\text{Ga}}$) which disrupts the Ga-Ga dimer bond, and (ii) In at the interstitial site In$_{\text{i}}$. Different charge states of the defects were investigated as a function of the Fermi energy. To identify the preferred location of the impurities, formation energies of the defects were calculated. All the atomic positions in the pure and defect containing supercells were fully relaxed until the quantum mechanical forces were less than 0.02 eV/\AA.
The formation energy of a defect $D$ in a charge state $q$ denoted as $D^q$ is given by~\cite{zhang;jpcm02}:
%\begin{widetext}
\begin{equation}
\label{eq.1}
\Delta H_{f}\left(D^{q}\right)=\Delta E\left(D^{q}\right)+\sum_{i}\left(n_{i}\mu_{i}\right)+qE_{F}
\end{equation}

%\begin{widetext}
\begin{equation}
\label{eq.2}
\Delta E\left(D^q\right)=E\left(D^q\right)-E\left(\text{GaSe}\right)+\sum_{i}\left(n_{i}E(i)\right)+q{{E}_{\text{VBM}}}
\end{equation}
%\end{widetext}

In eqs.~(\ref{eq.1}) and~(\ref{eq.2}), $E\left(D^q\right)$ and $E\left(\text{GaSe}\right)$ are the total energies of the defect-containing and the defect-free supercells and $E(i)$'s ($i =$ Ga, Se, defect) are the energies of the constituents in their standard solid state. The atomic chemical potentials $\mu_{i}$'s are referenced to $E(i)$ and $n_{i}$'s are the number of atoms removed from $(n_{i} > 0)$ or added to $(n_{i} < 0)$ the system. The electron chemical potential (Fermi energy) $E_{F}$ is referenced to the energy $(E_{\text{VBM}})$ of the valence band maximum (VBM). In the present calculations $(E_{\text{VBM}})$ was determined as the average of the one-electron energy level of the VBM over the $k$-points where the total energy was calculated. As pointed out by Zhang~\cite{zhang;jpcm02},  this approach has the advantage that the band edges calculated this way are consistent with the defect transition levels and gives a better single-particle position for the shallow defects. Furthermore, the band gap of GaSe calculated with this ``average band-edge'' approach $\left( \text{E}_{\text{gap}}^{\text{average}}=1.6\text{8}\ \text{eV} \right)$ is closer to the experimental value $\left( \text{E}_{\text{gap}}^{\text{exp}\text{.}}=2.1\text{3}\,\text{eV} \right)$ than the direct gap located at the $\Gamma-\text{point}$ $\left( \text{E}_{\text{gap}}^{\Gamma }=0.85\,\text{eV} \right)$.

To check the accuracy of our calculations vis-a-vis experiment we also calculated the charge transition level energies which correspond to the values of $E_{F}$   where the formation energies of a defect in two different charge states ($q$ and $q'$) are equal i.e.~\cite{zhang;jpcm02}:

\begin{equation}
\epsilon (q/{q}')={\left[ \Delta E(D^{q'})-\Delta E(D^q) \right]}/{\left( q-{q}' \right)}
\end{equation}

Equation~(\ref{eq.1}) shows that the formation energies of the defects depend on the atomic chemical potential $(\mu)$ of the constituents as well as on the charge state $(q)$ of the impurity. The values of the $\mu_{i}$'s are constrained by several physical conditions. For example: (a) to avoid precipitations, $\mu_{i}$'s must be negative and (b) to maintain a stable host compound, the chemical potentials must satisfy ${{\mu }_{\text{Ga}}}+{{\mu }_{\text{Se}}}=\Delta H\left( \text{GaSe} \right)$, where $\Delta H\left( \text{GaSe} \right)$ is the formation enthalpy of GaSe. The theoretical value of $\Delta H\left( \text{GaSe} \right)=-1.12\ \text{eV}$. To avoid secondary phase formation between the host elements and impurities, one must impose additional constraints on the chemical potentials of the defects~\cite{zhang;jpcm02}. However, the effect of these conditions would be a constant shift in the formation energies and since we are interested in the relative formation energies associated with different locations and charge states of the same defect we set ${{\mu }_{def}}={{\mu }_{\text{In}}}=0$.

In Fig.~\ref{fig;struct_form_ene}(b) we give the calculated formation energies as a function of $E_{F}$ for the two types of defects each in three different charge states. Also we give both theoretical $\left( \text{E}_{\text{gap}}^{\text{theor}}=1.68\,\text{eV} \right)$ and experimental $\left( \text{E}_{\text{gap}}^{\text{exp}}=2.13\,\text{eV} \right)$ band gaps. We find that he formation energy of $\text{In}_{\text{Ga}}^{0}$ is 0.28 eV and one of the charge transition level associated with this defect, $\epsilon (+1/0) = 0.28\,\text{eV}$ above the VBM. This value is in fairly good agreement with the acceptor level at 0.21 eV, measured by Cui et al. using deep level transient spectroscopy (DLTS)~\cite{cui;jap08}. This gives us confidence in our total energy calculations using DFT and the supercell model to understand the defect physics.
%
%%%%%%%%%%%%%%%%%%%%%%%%%%%%%%%%%%%%%%%%%%%%%%%%%%%%%%%%%%%%%%%%
%   FIGURE
%%%%%%%%%%%%%%%%%%%%%%%%%%%%%%%%%%%%%%%%%%%%%%%%%%%%%%%%%%%%%%%%

\begin{figure}[tbp]
\begin{center}$\,$
\begin{tabular}{c c}
\includegraphics[width=1.6875in,keepaspectratio=1]{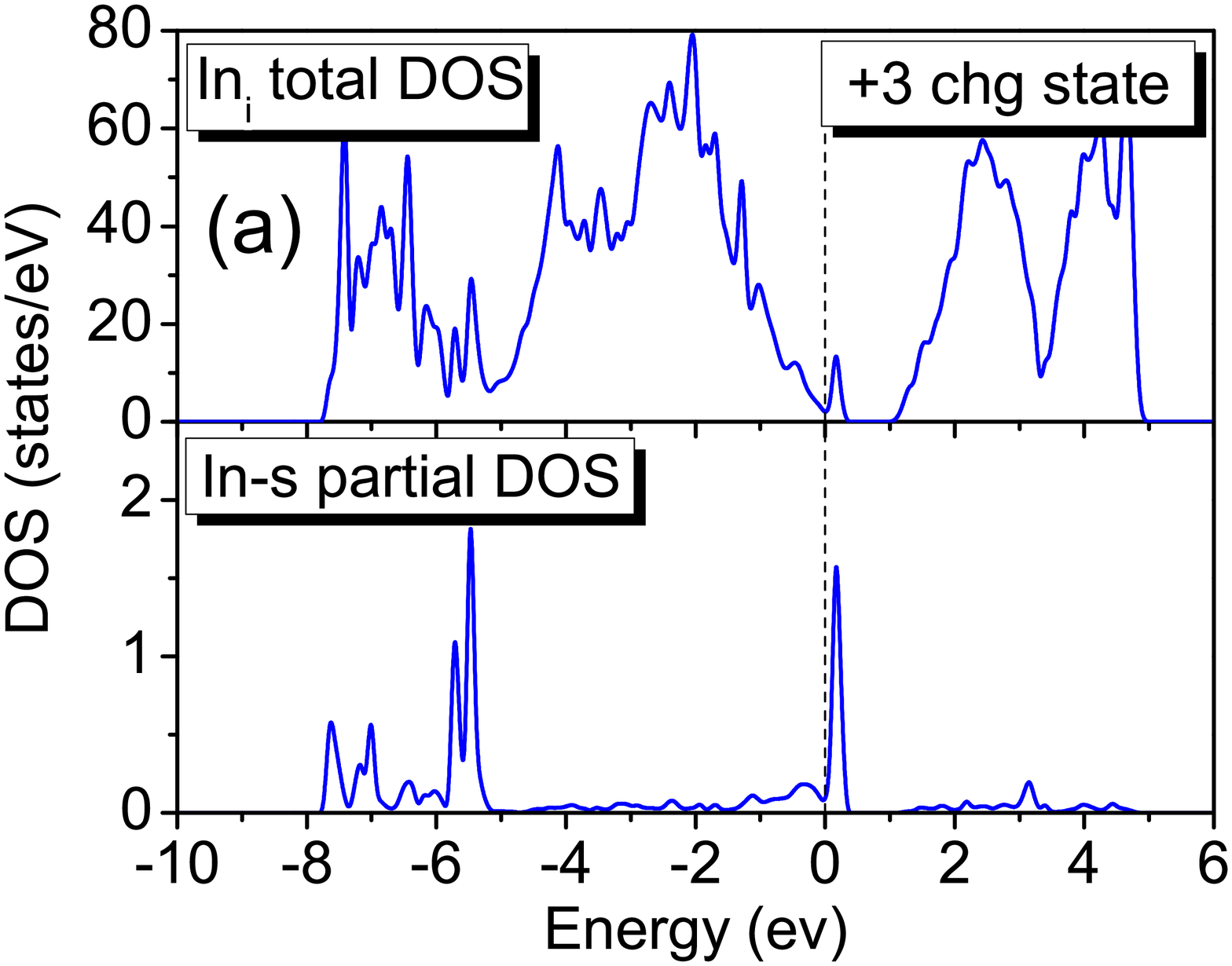} &
\includegraphics[width=1.6875in,keepaspectratio=1]{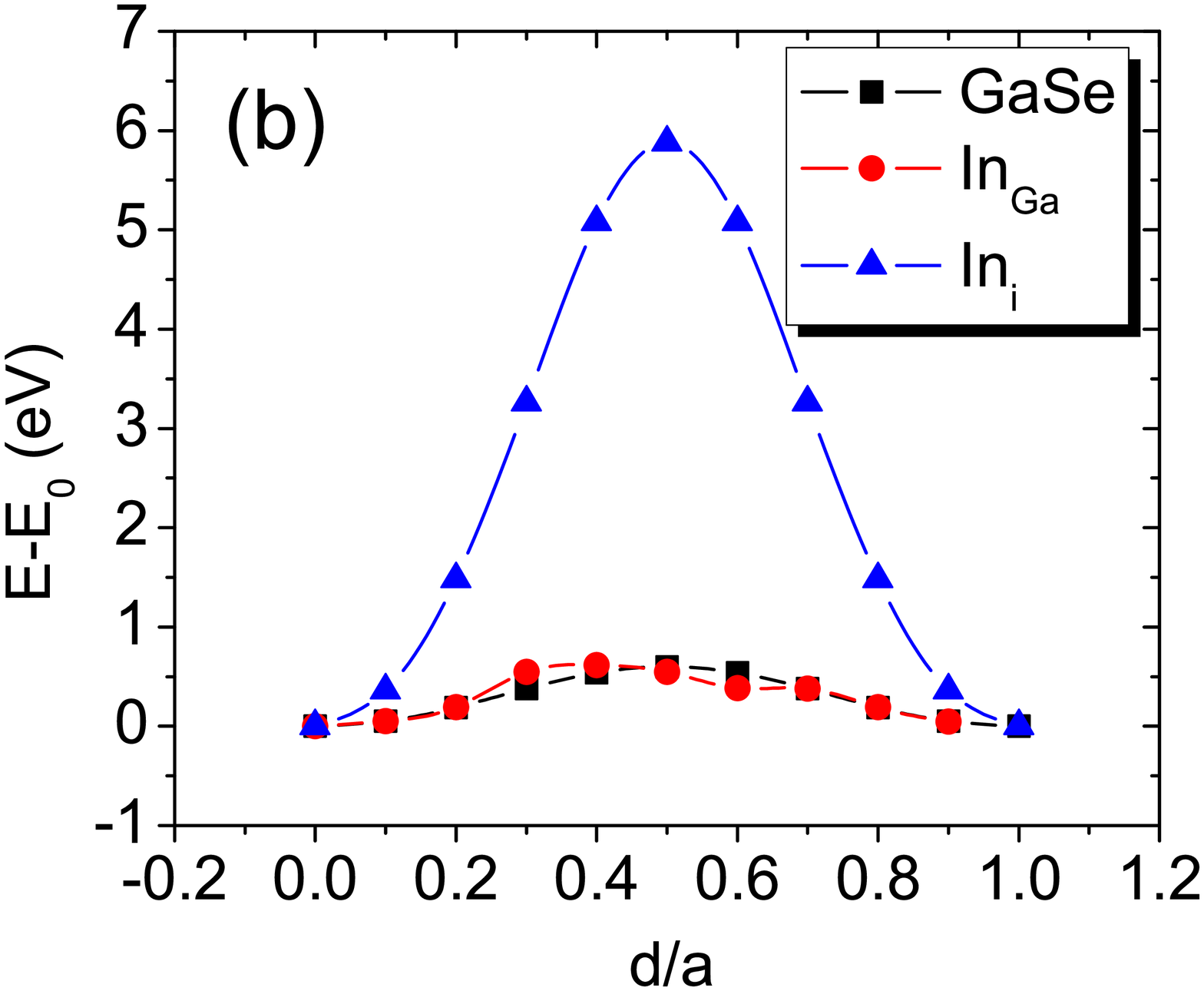} \\
\end{tabular}
\end{center}
\caption {(a) The total DOS of GaSe with In$_{i}$ and the projected DOS of the In $s$-orbital, showing the positions of the HDDS (-5.5 eV) and DDS (just above $E_{F}$) introduced by the charged In$_{\text{i}}$ defect; (b) The energy barrier which must be overcome in order to cleave the GaSe crystal increases dramatically when In occupies the interstitial site compared to the case when In occupies substitutional site. For comparison the case of pure GaSe is also shown.}
\protect\label{fig;dos_cleavage}
\end{figure}

The defect states with lowest formation energies are: $\text{In}_{\text{i}}^{\text{3+}}$ for ${{E}_{\text{VBM}}}(=0\,\text{eV)}\le {{E}_{F}}\le 0.135\,\text{eV}$, $\text{In}_{\text{Ga}}^{\text{1+}}$ for $0.135\,\text{eV}\le {{E}_{F}}\le 0.28\,\text{eV}$ and $\text{In}_{\text{Ga}}^{\text{0}}$ for $0.28\ \text{eV}\le {{E}_{F}}$. $\text{In}_{\text{Ga}}^{\text{0}}$ is the most stable defect for a wide range of $E_{F}$. However when the Fermi energy is tuned towards the VBM energy, $\text{In}_{\text{i}}^{\text{3+}}$ defect becomes most stable. We will discuss the underlying physics of this change by examining the single particle density of states (see below). As regards the effect of $\text{In}_{\text{Ga}}^{\text{0}}$ defect on the electronic structure, we find that the band structures near VBM and CBM are affected very little. One therefore does not expect much change in the transport properties in In doped GaSe if the impurity goes to a Ga site in the neutral charge state. One can understand this lack of significant change by looking at the $\text{In}\, 5s-\text{Ga}\,4s$ dimer antibonding state (which hybridizes with the Se $p$-bands to give rise to states in the neighborhood of the band gap) and observe that it is not significantly different from the $\text{Ga}\,4s-\text{Ga}\,4s$ dimer antibonding state.

To understand why $\text{In}_{\text{i}}^{\text{3+}}$ has the lowest formation energy when${{E}_{F}}\le 0.135\,\text{eV}$, we look at the electronic structure, the single particle density of states (DOS) and the nature of defect state introduced by In$_{\text{i}}$. Fig.~\ref{fig;dos_cleavage}(a) gives the total and partial (associated with In $s$) DOS for this case. We see that In$_{\text{i}}$ introduces a hyper deep defect state (HDDS) near the bottom of the Se $p$ bands (at $\sim -5.5\,\text{eV}$). It is a bonding state formed out of In $5s$ and neighboring Se $p$ states. The corresponding antibonding state splits off from the Se $p$ valence band states and is denoted as the deep defect state (DDS). This picture is very close to what happens when In is a substitutional defect in PbTe~\cite{ahmad;prl06}. The strong mixing between In $5s$ and the neighboring Se $p-$states leads to the removal of one state (per spin) from the Se $p$ band which becomes the DDS. In terms of electron counting, two of the three electrons from In occupy the HDDS and the three electrons (two from the electrons occupying the valence band in pure GaSe and one from In) fill the DDS and partially occupy the conduction band. Thus $\text{In}_{\text{i}}$ acts like a donor. Since the three electrons occupy states with energies larger than $\text{E}_{\text{VBM}}$, clearly the formation energy of $\text{In}_{\text{i}}$ in charge state $q$ = 0, 1, and 2 are higher than In$_{\text{Ga}}$ for which neither the band structure nor the electron count change. By removing three electrons from $\text{In}_{\text{i}}$ to obtain $q = 3$ charge state we can lower its formation energy~\cite{comment}.

Now that we understand the microscopic nature of the In defect states corresponding to two different positions we have to see whether the changes in the local bonding lead to appreciable modifications of the elastic stiffness of the GaSe matrix as seen experimentally~\cite{frene;pcgcm94,voevo;opmat04}. To simplify our analysis of elastic stiffness we have used two approaches, one by calculating the elastic constants and the other by exploring the energy barrier involved in relative shearing of two blocks in a super cell.  For In$_{\text{i}}$ we use the second approach whereas for $\text{In}_{\text{Ga}}$ $(\text{Ga}_{1-x}\text{In}_{x}\text{Se})$ we use both the methods. $\text{Ga}_{1-x}\text{In}_{x}\text{Se}$ is characterized by 5 elastic constants: $\text{C}_{11}$, $\text{C}_{12}$, $\text{C}_{13}$, $\text{C}_{33}$, and $\text{C}_{44}$. We have determined these quantities from total energy calculations for five different strain configurations~\cite{wrigh;jap97}. For each $x$ we have calculated the theoretical crystal structures by minimizing the total energies with respect to the lattice constants: first with respect to the volume of the unit cell keeping the $c/a$ ratio fixed and then with respect to $c/a$ keeping the previously obtained equilibrium volume constant. The elastic constants for $x$ = 0, 0.25, 1 were obtained using small unit cells (8 atoms/cell) and the Brillouin zone (BZ) was sampled by a $\Gamma-\text{centerd}$ 12x12x3 $k-$mesh. In the case of $x=0.0625$ the calculations were performed on 2x2x1 supercell with the BZ sampled by a 6x6x3 grid of $k-$points.

\begin{table*}%[H] add [H] placement to break table across pages
 \caption{Elastic constants of $\text{Ga}_{1-x}\text{In}_{x}\text{Se}$. All values are given in GPa.\label{table;elast_const}}
 %\begin{ruledtabular}{c l c c c c c c c }
 \begin{tabular}[t]{c @{\hspace{15pt}} l @{\hspace{15pt}} c @{\hspace{15pt}} c @{\hspace{15pt}} c @{\hspace{15pt}} c @{\hspace{15pt}}c @ {\hspace{15pt}} c @{\hspace{15pt}}c }
 \hline\hline
 $x$ & & $\text{C}_{11}$ & $\text{C}_{12}$ & $\text{C}_{13}$ & $\text{C}_{33}$ & $\text{C}_{44}$ & $\text{C}_{11}+\text{C}_{12}$ & $\text{C}_{11}-\text{C}_{12}$ \\
 \hline
 \multirow{3}{*} {0} & Present calc.                & 100.88 & 27.04 & 9.74 & 33.39 & 8.34 & 127.92 & 73.80  \\
                     & Ref.~\cite{adler;prb98}      &        &       & 12.70 & 35.40 &      & 130.20 &       \\
                     & Ref.~\cite{gatul;pss83}      & 105.00 & 32.4  & 12.60 & 35.10 & 10.40& 137.40 & 72.6  \\
 0.0625 &                                           & 99.17  & 27.12 & 10.47 & 34.37 & 8.79 & 126.29 & 72.05 \\
 0.25   &                                           & 91.28  & 25.63 & 11.34 & 36.05 & 9.67 & 116.91 & 65.65 \\
 \multirow{2}{*} {1} & Present calc.                & 70.34  & 23.51 & 14.19 & 38.49 & 11.53& 93.96  & 46.83 \\
                     & Ref.~\cite{gatul;pss83}      & 73.00  & 27.00 &       & 36.00 &      & 100.00 & 46.00 \\
 \hline\hline
 \end{tabular}
 %\end{ruledtabular}
 \end{table*}

We observe a monotonic increase in the lattice constants as the concentration of the impurities ($x$-value) increases following the usual Vegard's law. For the end compounds GaSe and InSe the theoretical lattice constants are less than 3\% smaller compared to experiment, while the $c/a$ ratios are within 0.7\% of the experimental values. The underestimation of the lattice parameters is due to the well-known overbinding effect of LDA. The calculated elastic constants are given in Table~\ref{table;elast_const}, along with previously calculated theoretical  results~\cite{adler;prb98} and experimental data~\cite{gatul;pss83}. There is a good overall agreement with the available experimental and theoretical values. The changes in the elastic constants of $(\text{Ga}_{1-x}\text{In}_{x}\text{Se})$ depend almost linearly on $x$: on one hand, we observe a monotonic decrease in $\text{C}_{11}$ and $\text{C}_{12}$, but on the other hand, the elastic constants $\text{C}_{13}$, $\text{C}_{33}$, and $\text{C}_{44}$ show small enhancement with increasing $x$. This indicates that when In impurity occupies Ga sites, the crystal becomes softer in the $a-$ and $b-$directions (parallel to the atomic layers) and stiffer along the $c-$axis (perpendicular to the atomic layers).

The effect of substitutional In on the elastic constants of GaSe can be understood if we examine the connection between the structural and electronic properties of the end compounds: GaSe and InSe. In the case of InSe the interlayer separation is ~2.96 \AA, which is smaller compared to the interlayer distance in GaSe (~3.15 \AA), indicating a stronger InSe interlayer interaction. Since the elasticity of the in the $c-$direction is determined predominantly by the interaction between the atomic layers, the crystal becomes stiffer with the increase of In concentration in GaSe. The softening of $(\text{Ga}_{1-x}\text{In}_{x}\text{Se})$ in the $a-$ and $b-$directions with the increase in the composition $x$, can be easily understood, because the intralayer distances are longer and therefore the intralayer covalent bonds are weaker in InSe than in GaSe~\cite{erran;prb05,camar;prb02}. Given that in the a- and b-directions there are no ``interlayer regions'' which could counteract the weakening of the atomic bonds, the crystal becomes softer as the In concentration increases. Although the substitutional In impurity seems to increase the elastic stiffness of GaSe along $c-$axis the, the effect is rather small (e.g. $\text{C}_{33}$ increases by 7.8\% from GaSe to InSe). So we have to look at a different mechanism for In induced increased inter-layer rigidity as seen experimentally.

As we see the elastic constants do not change appreciably in In doped systems when In goes as a substitutional impurity. This is also seen in the calculations of energy barrier associated with relative shearing of two atomic blocks (each block being made up from 4-atomic planes) in a unit cell. Fig.~\ref{fig;dos_cleavage}(b) compares the energy barriers involved in this relative shearing. The energy barriers for both pure GaSe and $\text{Ga}_{1-x}\text{In}_{x}\text{Se}$ are very small and comparable. Thus substitutional In does not enhance the shear rigidity of GaSe. In the same figure we show the energy barrier associated with similar shearing in the presence of an interstitial charged In defect ($\text{In}_{\text{i}}^{\text{3+}}$). The energy barrier and the initial slope increase dramatically (by factors of $\sim 10$ and $\sim 7$ respectively) in the presence of $\text{In}_{\text{i}}^{\text{3+}}$. Clearly GaSe is very soft and inserting interlayer charged In defects can make the crystal rigid against shear distortion. We note that for certain shear configurations the interstitial In could not be accommodated by atomic relaxation. Thus, for consistency, we present the results obtained for rigid shear.

In summary, using ab initio electronic structure studies we have shown that In defects can go into GaSe in three ways, (i) mostly as a neutral substitutional defect, (ii) as a singly charged substitutional defect and (iii) as a triply charged interstitial defect $\text{In}_{\text{i}}^{\text{3+}}$. The charged defects can be stabilized by p-doping GaSe.  As a substitutional defect In does not impact appreciably the physical properties of the crystal. In contrast the interstitial In if in a proper charge state can increase the shear stiffness of GaSe~\cite{voevo;opmat04} significantly. We suggest that the observed dramatic improvement when In is added to p-type GaSe is due to charged interstitial defects. One of the important predictions of our calculation is the presence of a deep defect state just above the valence band maximum when In is present as an interstitial defect. Deep level spectroscopy should be able to see this defect.


\begin{references}
\bibitem{comment} The formation energy of neutral In interstitial defect is underestimated in our calculation since on electron/defect occupies the bottom of the conduction band and the LDA gap is 0.85 eV compared to the average band gap of 1.68 eV and the experimental value of 2.13 eV.
\end{references}
\end{document}